\documentclass[%
 reprint,
 amsmath,amssymb,
 aps,
]{revtex4-2}

\usepackage{graphicx}
\usepackage{dcolumn}
\usepackage{bm}

\begin{document}

\preprint{APS/123-QED}

\title{A control of threshold of hard excitaion mode of optomechanical system by a low-intensity seed wave}

\author{Artem R. Mukhamedyanov}
\affiliation{Moscow Institute of Physics and Technology, 141700, 9 Institutskiy pereulok, Moscow, Russia}
\affiliation{Dukhov Research Institute of Automatics (VNIIA), 127055, 22 Sushchevskaya, Moscow, Russia}
\author{Alexander A. Zyablovsky}
\email{zyablovskiy@mail.ru}
\affiliation{Moscow Institute of Physics and Technology, 141700, 9 Institutskiy pereulok, Moscow, Russia}
\affiliation{Dukhov Research Institute of Automatics (VNIIA), 127055, 22 Sushchevskaya, Moscow, Russia}
\affiliation{Institute for Theoretical and Applied Electromagnetics, 125412, 13 Izhorskaya, Moscow, Russia}
\author{Evgeny S. Andrianov}
\affiliation{Moscow Institute of Physics and Technology, 141700, 9 Institutskiy pereulok, Moscow, Russia}
\affiliation{Dukhov Research Institute of Automatics (VNIIA), 127055, 22 Sushchevskaya, Moscow, Russia}
\affiliation{Institute for Theoretical and Applied Electromagnetics, 125412, 13 Izhorskaya, Moscow, Russia}

\date{\today}

\begin{abstract}
An optomechanical system based on two optical modes, one of which is pumped by a coherent wave, interacting with each other via a phonon mode is considered. In such a system, a hard excitation mode can be realized in the region of bistability. It is demonstrated that excitation of an optical mode with a lower frequency by a low-intensity seed wave can lead to a significant reduction of a threshold of the hard excitation mode. This decrease is due to the fact that the seed wave changes the stability of solutions in the bistability region. This leads to the fact that switching from a non-generating state to a generating one occurs at lower values of the intensity of the pump wave. The obtained result paves the way for the creation of all-optical transistors and logical elements.
\end{abstract}

\maketitle

Optomechanical systems, in which photon and phonon modes interact with each other, have wide applications in various areas such as high-precision measurement \cite{1, 2, 3}, ultrasound sensing \cite{4, 5, 6}, and quantum information processing \cite{7, 8}. During the past decade, significant progress has been made in both fundamental research and practical applications of such systems. Optomechanical systems have been used to demonstrate a variety of phenomena, such as phonon generation \cite{9, 10, 11}, collision of eigenvalues at exceptional points (EP) \cite{12, subthreshold}, optomechanically induced transparency \cite{14, 15}, etc. In view of the creation of Brillouin and phonon lasers, of particular interest is overcoming the threshold of optomechanical instability and achieving a self-oscillating regime of optical and phonon modes.

Recently, the existence of bistability has been demonstrated in an optomechanical system consisting of two optical modes, which interact with each other via a phonon mode \cite{hard_exc}. The external pump wave is used to excite an optical mode with a higher frequency. In such a system, there exists a region of parameters in which the states with low and high intensities are stable simultaneously \cite{hard_exc}. At a certain value of the pump wave amplitude, the system switches from the states with low intensity to the one with high intensity. Such switching leads to an abrupt increase in the intensity of the optical and phonon modes and corresponds to a hard excitation mode of laser generation. The hard excitation mode can be used, for example, for the creation of a new type of high-sensitivity sensors \cite{16, 17, hard_exc}, optical transistors \cite{19, 20}, and ultrafast optical switches \cite{21, 22}.

To achieve the hard excitation mode, it is necessary to ensure a certain frequency detuning between the pump wave and the pumped mode. Moreover, the higher the detuning, the higher the intensity jump at the hard excitation mode threshold. However, the high detuning results in the threshold of the hard excitation mode turning out to be an order of magnitude higher than the threshold of optomechanical instability. Therefore, reducing the threshold of the hard excitation mode is a critical task in order to minimize the pump energy consumption.

In this letter, we propose an approach to reduce the threshold of the hard excitation mode. We consider the optomechanical system pumped by two external waves. The first external wave, the pump wave, off-resonantly excites the optical mode with the higher frequency. We operate in the hard excitation mode, where the intensities of the optical modes and phonon mode jump near the threshold. We use a second external wave, the seed wave, to resonantly excite the lower-frequency optical mode. We demonstrate that the weak intensity of the seed wave leads to a noticeable decrease in the threshold of the hard excitation regime such that the total intensity of the two waves is of the order of magnitude smaller than the threshold intensity in a single-wave excitation scheme. This decrease is due to the fact that the seed wave changes the stability of solutions in the bistability region, which causes the system to switch from the low-intensity state to the high-intensity state at lower pump wave intensities. We show that in the soft excitation mode, due to the absence of a bistability region, the seed wave leads to only minor changes in the output-input curves.

We consider an optomechanical system consisting of two optical modes interacting with each other via the phonon mode. The frequencies of the optical modes are $\omega_{1,2}$, $\omega_1 > \omega_2$. We suppose that they differs from each other by the frequency of the phonon mode, $\omega_b$ (i.e., $\omega_1 - \omega_2 = \omega_b$). The pump and seed waves excite the optical modes with the frequencies $\omega_1$ and $\omega_2$, respectively. To describe the system, we use the optomechanical Hamiltonian \cite{1}:

\begin{equation}
\begin{gathered}
  \hat H = \hbar {\omega _1}\hat a_1^\dag {{\hat a}_1} + \hbar {\omega _2}\hat a_2^\dag {{\hat a}_2} + \hbar {\omega _b}{{\hat b}^\dag }\hat b +  \hfill \\
  \hbar \,g\left( {\hat a_1^\dag {{\hat a}_2}\hat b + {{\hat a}_1}\hat a_2^\dag {{\hat b}^\dag }} \right) + \hbar \Omega_{1} \left( {\hat a_1^\dag {e^{ - i\omega t}} + {{\hat a}_1}{e^{i\omega t}}} \right) \hfill + \\ \hbar \Omega_{2} \left( { \,\hat a_2^\dag {e^{ - i{\omega _{2}}\,t}} + {{\hat a}_2}{e^{i{\omega _{2}}\,t}}} \right)
\end{gathered}
\label{eq:1}
\end{equation}
Here ${\hat a_{1,2}}$ and $\hat a_{1,2}^\dag $ are the bosonic annihilation and creation operators for the first and the second optical modes, respectively. $\hat b$ and ${\hat b^\dag}$ are the annihilation and creation operators of the phonon mode. $g$ is the coupling strength between the optical modes and the phonon mode (Frohlich constant). $\Omega_{1,2}$ are the amplitudes of the pump and seed waves, respectively. $\omega$ is the frequency of the pump wave. Here, the second optical mode and the seed wave are in resonance. The Hamiltonian~(\ref{eq:1}) describes the optomechanical system under the rotating-wave approximation, where the rapidly oscillating terms are not taken into account \cite{allen}. This approximation remains valid under the condition that the coupling strength is significantly smaller than the oscillation frequencies of the system \cite{allen}, that is, when $g \ll {\omega _{1,2}}$.

The interaction of the system under consideration with the environment leads to energy relaxation \cite{carmichael,gardiner}. To describe relaxation processes, we use the Langevin approach \cite{39}. Within this framework, we derive the equations for the c-number amplitudes of the optical and phonon modes that have the form~\cite{hard_exc}:

\begin{equation}
\frac{{d{a_1}}}{{dt}} = \left( { - {\gamma _1} - i{\omega _1}} \right){a_1} - ig{a_2}b - i\Omega_{1} {e^{ - i{\omega}t}}
\label{eq:2}
\end{equation}

\begin{equation}
\frac{{d{a_2}}}{{dt}} = \left( { - {\gamma _2} - i{\omega _2}} \right){a_2} - i{g}{a_1}{b^*} - i\Omega_{2} {e^{ - i{\omega _{2}}\,t}}
\label{eq:3}
\end{equation}

\begin{equation}
\frac{{db}}{{dt}} = \left( { - {\gamma _b} - i{\omega _b}} \right)b - i{g}{a_1}a_2^*
\label{eq:4}
\end{equation}
Here ${a_{1,2}}$ and $b$ are the c-number amplitudes of the optical modes and phonon mode, respectively. ${\gamma _{1,2}}$, ${\gamma _b}$ are the relaxation rates of the respective quantities. In our calculations, we consider that both the pump and seed waves are included at the initial moment of time, $t =0 $. Before the initial moment of time, the amplitudes of the external wave are zero ($\Omega_1 = \Omega_2 = 0$) and the system is in an unexcited state ($a_1 = a_2 = b = 0$).

We study the behavior of the optomechanical system for various values of the system parameters and the amplitudes of the pump and the seed waves. The numerical simulation of the Eqns.~(\ref{eq:2})-(\ref{eq:4}) shows that there are stationary states of the Eqns.~(\ref{eq:2})-(\ref{eq:4}), for which the intensities of the optical modes and the phonon mode do not depend on time. To derive the expressions for the stationary states, we make the following substitutions: ${a_1} = {a_{1st}}{e^{ - i{\omega}\,t}}$, ${a_2} = {a_{2st}}{e^{ - i{{\omega _{2}} }t}}$, $b = {b_{st}}{e^{-i{\left(\omega - \omega _{2}\right)}t}}$, and introduce the frequencies detunings $\delta \omega_{1,2} = \omega_{1,2} - \omega_{}$, $\Delta_2 = 0$, $\delta \omega_b = \omega_b + \delta\omega_{2}$ \cite{hard_exc}. As a result, we obtain the following equations for the stationary amplitudes of the modes:

\begin{equation}
0 = \left( { - {\gamma _1} - i{\delta \omega _1}} \right){a_{1st}} - ig{a_{2st}}b_{st} - i\Omega_{1}
\label{eq:5}
\end{equation}

\begin{equation}
0 = \left( { - {\gamma _2} - i{\Delta_2}} \right){a_{2st}} - i{g}{a_{1st}}{b_{st}^*} - i\Omega_{2}
\label{eq:6}
\end{equation}

\begin{equation}
0 = \left( { - {\gamma _b} - i{\delta\omega _b}} \right)b_{st} - i{g}{a_{1st}}a_{2st}^*
\label{eq:7}
\end{equation}

\begin{figure*}[htbp]
\centering\includegraphics[width=\linewidth]{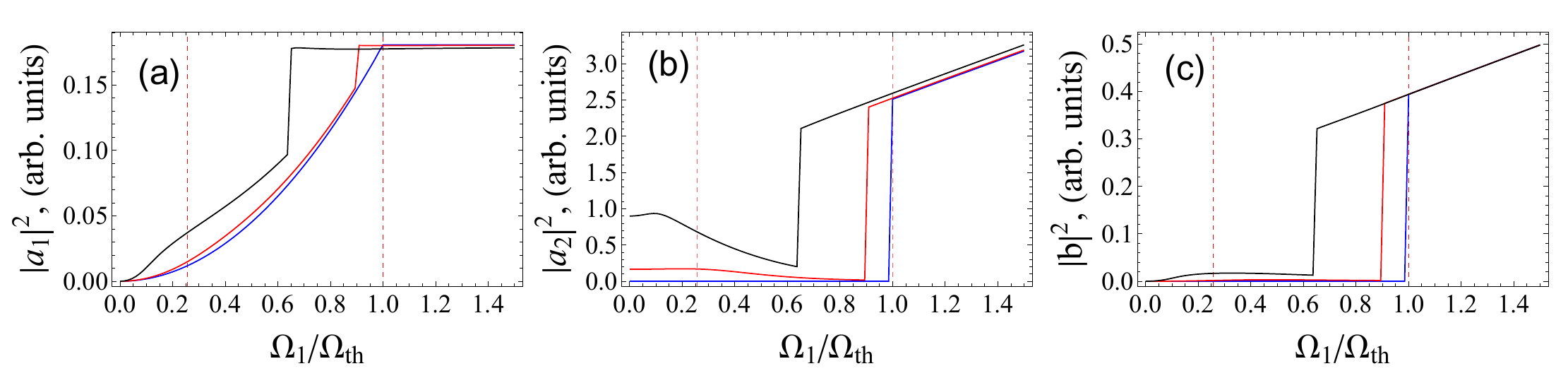}
\caption{Dependence of the intensity of the first optical mode (a), the second optical mode (b) and the phonon mode (c) on the pump amplitude of the pump wave, $\Omega_1$, when the amplitude of the seed wave is $\Omega_2 = 0$ (the solid blue lines); $\Omega_2 = 3\cdot10^{-2} \Omega_{th}$ (the solid red lines); $\Omega_2 = 7\cdot 10^{-2} \Omega_{th}$ (the solid black lines). The vertical dashed lines show $\Omega_{ex}$ and $\Omega_{th}$, respectively. Here $\Omega_2 = 7\cdot 10^{-2} \Omega_{th}$; $\gamma_1/2\pi = \gamma_2 / 2\pi = 19$ MHz, $\gamma_b/2\pi = 121$ MHz, ${\delta\omega}_1/2\pi = \delta \omega_b/2\pi = -6$ GHz, $\Delta_2/2\pi = 0$ GHz, $g/2\pi = 1.6$ kHz, $\Omega_{th} = 5 \cdot 10^{14}$ s$^{-1}$. These parameters correspond to the hard excitation mode of generation when $\Omega_2 = 0$. For all values of the amplitudes of the pump and seed waves, the initial state is $a_1 = a_2 = b = 0$.}
\label{fig:2}
\end{figure*}

\begin{figure*}[htbp]
\centering\includegraphics[width=\linewidth]{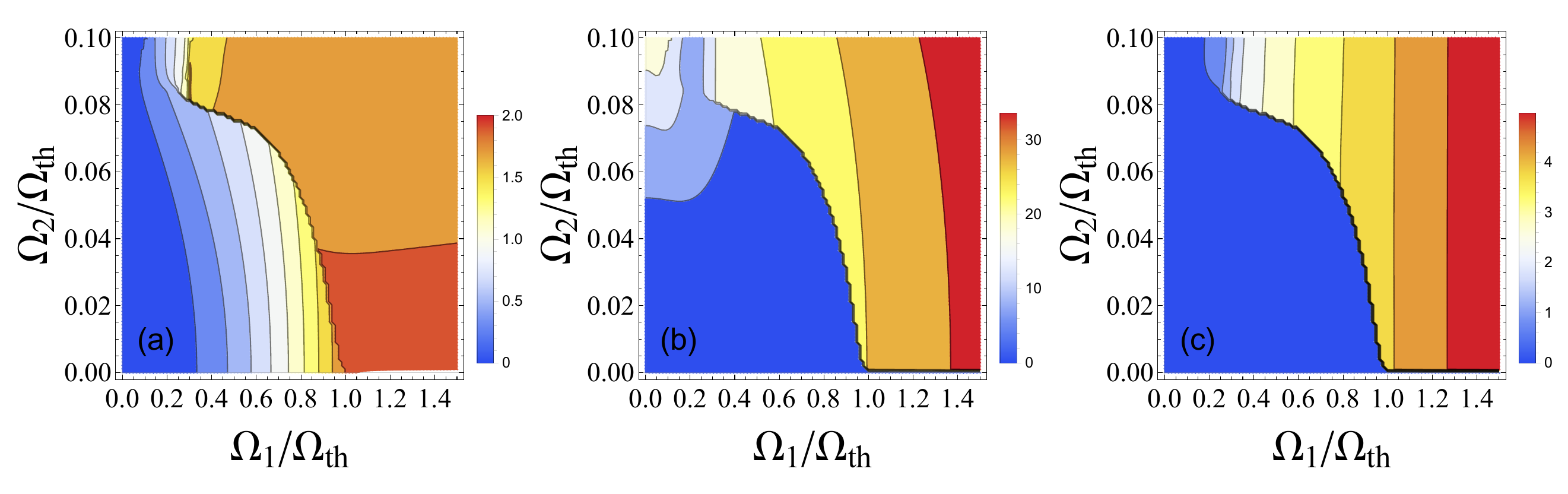}
\caption{Dependence of the intensity of the first optical mode (a), the second optical mode (b) and the phonon mode (c) on the amplitudes of the pump $\Omega_1$ and the seed $\Omega_2$ waves. All other parameters are the same as in Figure~\ref{fig:2}.}
\label{fig:3}
\end{figure*}

In the general case, the Eqns.~(\ref{eq:5})-(\ref{eq:7}) have not analytical solution and can be solved only in limiting cases such as $\Omega_2 = 0$ or $\Omega_1 = 0$. When $\Omega_2 = 0$ and $\Omega_1 \ne 0$, it has been demonstrated in \cite{hard_exc} that there are two stationary solutions: a zero solution and a non-zero solution. For the zero solution, the amplitudes of the second optical mode and the phonon mode are zero, and the amplitude of the first mode is proportional to $\Omega_1$. The stability of solutions depends on the value of $\Omega_1$ and the values of the system parameters. There is a bistable region in which both solutions are simultaneously stable \cite{hard_exc}. When the pump wave intensity exceeds a certain value - a threshold of the hard excitation mode - switching between solutions occurs. It leads to an abrupt increase in the intensities of the second optical mode and the phonon mode~\cite{hard_exc} [Figure~\ref{fig:3}]. The hard excitation mode takes place when $\delta \omega_1 \left( \delta \omega_2 + \omega_b \right) > \gamma_1 \left(\gamma_2 + \gamma_b \right)$~\cite{hard_exc}. When this inequality is satisfied, the area of the bistable region is $\Omega_{ex} \leqslant |\Omega_1| \leqslant \Omega_{th}$, where $\Omega_{ex} = \frac{\sqrt{\gamma_b \gamma_2}}{|g|} \left | {\delta \omega}_1 +  \frac{{\delta \omega}_2 + \omega_b}{\gamma_2 + \gamma_b} \gamma_1 \right|$ and $\Omega_{th} = \frac{1}{|g|}\frac{\sqrt{\gamma_b \gamma_2}}{\gamma_2 + \gamma_b}\sqrt{\left(\gamma_1^2 + {\delta\omega}_1^2 \right) \left(1 + \left(\frac{{\delta \omega}_2 + \omega_b}{\gamma_2 + \gamma_b}\right)^2 \right)}$~\cite{hard_exc}.

In the general case, when $\Omega_1 \ne 0$ and $\Omega_2 \ne 0$, the amplitudes of the modes can be calculated by the numerical simulation of the Eqns.~(\ref{eq:2})-(\ref{eq:4}). The numerical simulation shows that when $\Omega_2 \ne 0$, the intensities of the optical and phonon modes are not zero for all values of $\Omega_1$ [Figure~\ref{fig:2}]. There are two solutions: the low-intensity solution, which goes into the zero solution at $\Omega_2 = 0$, and the high-intensity solution. Moreover, there is a bistable region in which both solutions are stable. An increase in the amplitude of $\Omega_1$ leads to the abrupt increase in the mode intensities, which, by analogy with the case when $\Omega_2 = 0$. That is, the hard excitation mode can take place in the general case. It is interesting that before the threshold of the hard excitation mode, the intensities of the modes can change non-monotonically with increasing $\Omega_1$ [Figure~\ref{fig:2}]. Such a non-monotonicity is associated with a change in the frequencies at which the maximum response of the system to external pumping is achieved.

\begin{figure}[htbp]
\includegraphics[width=\linewidth]{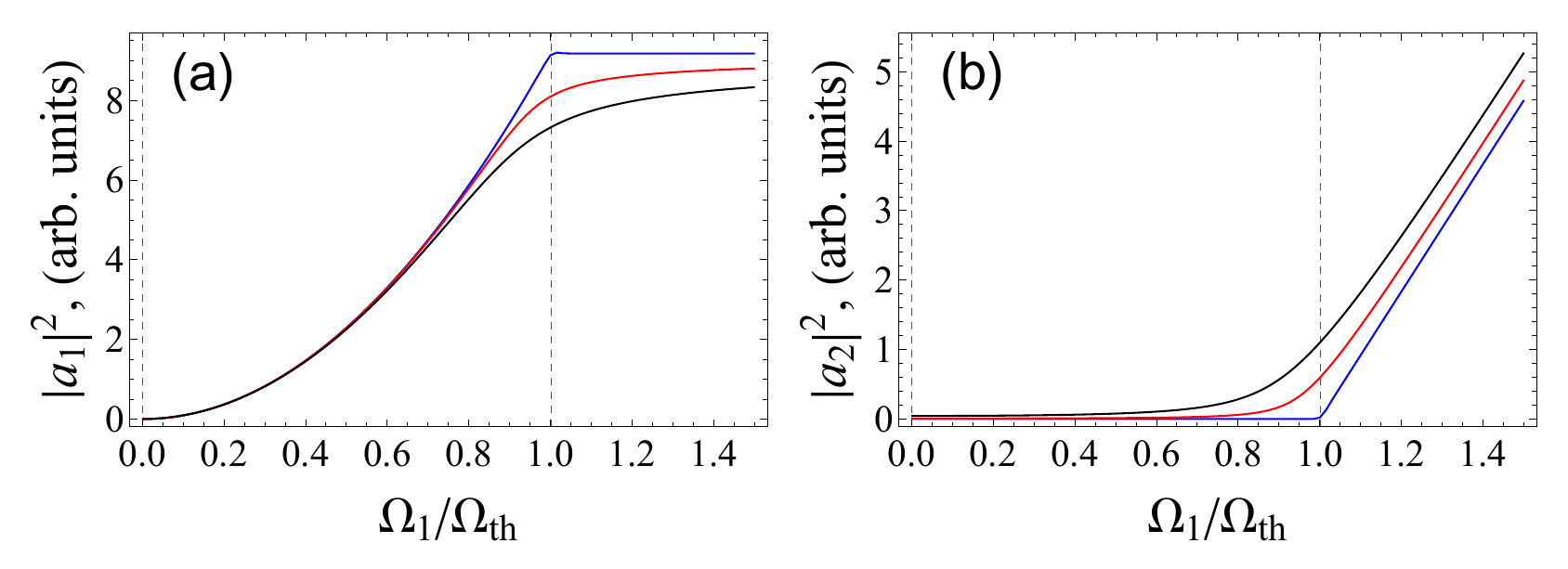}
\caption{Dependence of the intensity of the first optical mode (a), the second optical mode (b) on the pump amplitude of the pump wave $\Omega_1$ when $\Omega_2 = 0$ (the solid blue lines); $\Omega_2 = 3\cdot10^{-2} \Omega_{th}$ (the solid red lines); $\Omega_2 = 7 \cdot 10^{-2} \Omega_{th}$ (the solid black lines). The vertical dashed line shows $\Omega_{th}$. Here $\Omega_{th} = 3.6\cdot10^{-12}$ s$^{-1}$. All other parameters are the same as Figure \ref{fig:2}. These parameters correspond to the soft excitation mode of generation when $\Omega_2 = 0$.}
\label{fig:4}
\end{figure}

We calculate the dependencies of the modes' intensities and the threshold of the hard excitation mode on $\Omega_1$ and $\Omega_2$ [Figure~\ref{fig:3}]. Our calculations show that an increase in $\Omega_2$ leads to a decrease in the hard excitation mode threshold from $\Omega_{th}$ to a value $\Omega_{ex}$ [Figure~\ref{fig:3}]. The decrease in the threshold takes place in the bistable region and, therefore, relates to the co-existence of the low-intensity and the high-intensity solutions. 

\begin{figure}[htbp]
\includegraphics[width=\linewidth]{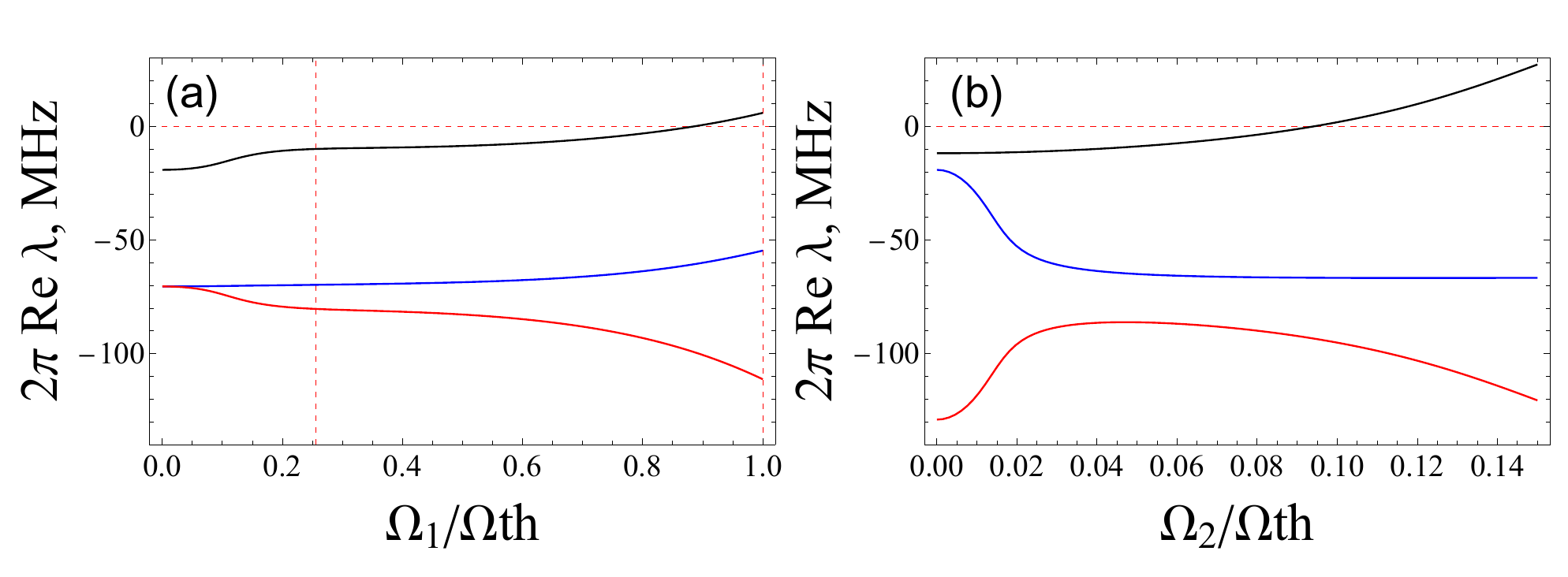}
\caption{Dependence of the real parts of the eigenvalues of the linearized system for the small deviation from the non-generating solution on $\Omega_1$ (a) and $\Omega_2$ (b). In the case (a) $\Omega_2 = 7\cdot 10^{-2} \Omega_{th}$. The vertical dashed lines correspond to $\Omega_{ex}$ and $\Omega_{th}$, respectively. All other parameters are the same as in Figure~\ref{fig:2}. In the case (b) $\Omega_1 = 7\cdot 10^{-1} \Omega_{th}$.}
\label{fig:5}
\end{figure}

Based on the linear stability analysis of the low-intensity solution, we find that such a solution becomes less stable at $\Omega = \Omega_{ex}$. That is, the absolute value of the real part of the least stable eigenvalue decreases [see the black line in Figure~\ref{fig:5}a]. In other words, the appearance of the high-intensity solution decreases the stability of the low-intensity solution. In addition, the increase in $\Omega_2$ also decreases the stability of the low-intensity solution [Figure~\ref{fig:5}b]. In the bistable region, the decrease in the stability of the low-intensity solution causes the system to switch to the high-intensity state at lower values of $\Omega_1$. This corresponds to a decrease in the threshold of the hard excitation mode [Figures~\ref{fig:2} and \ref{fig:3}].

Thus, we can conclude that the change in the stability of the low-intensity solution caused by the seed wave can lead to the decrease in the threshold of the hard excitation mode. It is important that in the case of the soft excitation mode, the increase in the amplitude of the seed wave leads to an insignificant change in the intensity of the modes [Figure~\ref{fig:4}]. This is because in this case there is no bistability region, and the decrease in the stability of the low-intensity solution cannot lead to a switch to another stable solution. In this case, the wave can only change the stability boundary of the solutions (the generation threshold), which changes slightly due to the low intensity of the seed wave [Figure~\ref{fig:4}]. Our calculations show that in the hard excitation mode, the use of the seed wave makes it possible to reduce the total pump intensity of the pump and seed waves required for the transition to the high-intensity solution by several times. 

In summary, we propose an approach to reduce the threshold of the hard excitation mode in the optomechanical system. We demonstrate that using the pump and seed waves makes it possible to reduce the total pump wave intensity required for generation compared to using a single pump wave. We argue that the seed wave leads to a change in the stability of the low-intensity state, which, in turn, simplifies the transition to the high-intensity solution, which is accompanied by a jump-like change in the intensity of the modes. Our results may be useful for the development of a new generation of optical and optomechanical devices.

\section*{Acknowledgment}
A.R.M. and E.S.A. thank the foundation for the advancement of theoretical physics and mathematics “Basis”.

\nocite{*}

\bibliography{apssamp}

\providecommand{\noopsort}[1]{}\providecommand{\singleletter}[1]{#1}%
\begin{thebibliography}{26}%
\makeatletter
\providecommand \@ifxundefined [1]{%
 \@ifx{#1\undefined}
}%
\providecommand \@ifnum [1]{%
 \ifnum #1\expandafter \@firstoftwo
 \else \expandafter \@secondoftwo
 \fi
}%
\providecommand \@ifx [1]{%
 \ifx #1\expandafter \@firstoftwo
 \else \expandafter \@secondoftwo
 \fi
}%
\providecommand \natexlab [1]{#1}%
\providecommand \enquote  [1]{``#1''}%
\providecommand \bibnamefont  [1]{#1}%
\providecommand \bibfnamefont [1]{#1}%
\providecommand \citenamefont [1]{#1}%
\providecommand \href@noop [0]{\@secondoftwo}%
\providecommand \href [0]{\begingroup \@sanitize@url \@href}%
\providecommand \@href[1]{\@@startlink{#1}\@@href}%
\providecommand \@@href[1]{\endgroup#1\@@endlink}%
\providecommand \@sanitize@url [0]{\catcode `\\12\catcode `\$12\catcode `\&12\catcode `\#12\catcode `\^12\catcode `\_12\catcode `\%12\relax}%
\providecommand \@@startlink[1]{}%
\providecommand \@@endlink[0]{}%
\providecommand \url  [0]{\begingroup\@sanitize@url \@url }%
\providecommand \@url [1]{\endgroup\@href {#1}{\urlprefix }}%
\providecommand \urlprefix  [0]{URL }%
\providecommand \Eprint [0]{\href }%
\providecommand \doibase [0]{https://doi.org/}%
\providecommand \selectlanguage [0]{\@gobble}%
\providecommand \bibinfo  [0]{\@secondoftwo}%
\providecommand \bibfield  [0]{\@secondoftwo}%
\providecommand \translation [1]{[#1]}%
\providecommand \BibitemOpen [0]{}%
\providecommand \bibitemStop [0]{}%
\providecommand \bibitemNoStop [0]{.\EOS\space}%
\providecommand \EOS [0]{\spacefactor3000\relax}%
\providecommand \BibitemShut  [1]{\csname bibitem#1\endcsname}%
\let\auto@bib@innerbib\@empty
\bibitem [{\citenamefont {Gavartin}\ \emph {et~al.}(2012)\citenamefont {Gavartin}, \citenamefont {Verlot},\ and\ \citenamefont {Kippenberg}}]{1}%
  \BibitemOpen
  \bibfield  {author} {\bibinfo {author} {\bibfnamefont {E.}~\bibnamefont {Gavartin}}, \bibinfo {author} {\bibfnamefont {P.}~\bibnamefont {Verlot}},\ and\ \bibinfo {author} {\bibfnamefont {T.~J.}\ \bibnamefont {Kippenberg}},\ }\bibfield  {title} {\bibinfo {title} {A hybrid on-chip optomechanical transducer for ultrasensitive force measurements},\ }\href@noop {} {\bibfield  {journal} {\bibinfo  {journal} {Nat. Nanotechnol.}\ }\textbf {\bibinfo {volume} {7}},\ \bibinfo {pages} {509} (\bibinfo {year} {2012})}\BibitemShut {NoStop}%
\bibitem [{\citenamefont {Krause}\ \emph {et~al.}(2012)\citenamefont {Krause}, \citenamefont {Winger}, \citenamefont {Blasius}, \citenamefont {Lin},\ and\ \citenamefont {Painter}}]{2}%
  \BibitemOpen
  \bibfield  {author} {\bibinfo {author} {\bibfnamefont {A.~G.}\ \bibnamefont {Krause}}, \bibinfo {author} {\bibfnamefont {M.}~\bibnamefont {Winger}}, \bibinfo {author} {\bibfnamefont {T.~D.}\ \bibnamefont {Blasius}}, \bibinfo {author} {\bibfnamefont {Q.}~\bibnamefont {Lin}},\ and\ \bibinfo {author} {\bibfnamefont {O.}~\bibnamefont {Painter}},\ }\bibfield  {title} {\bibinfo {title} {A high-resolution microchip optomechanical accelerometer},\ }\href@noop {} {\bibfield  {journal} {\bibinfo  {journal} {Nat. Photonics}\ }\textbf {\bibinfo {volume} {6}},\ \bibinfo {pages} {768} (\bibinfo {year} {2012})}\BibitemShut {NoStop}%
\bibitem [{\citenamefont {Xiong}\ \emph {et~al.}(2017)\citenamefont {Xiong}, \citenamefont {Si},\ and\ \citenamefont {Wu}}]{3}%
  \BibitemOpen
  \bibfield  {author} {\bibinfo {author} {\bibfnamefont {H.}~\bibnamefont {Xiong}}, \bibinfo {author} {\bibfnamefont {L.-G.}\ \bibnamefont {Si}},\ and\ \bibinfo {author} {\bibfnamefont {Y.}~\bibnamefont {Wu}},\ }\bibfield  {title} {\bibinfo {title} {Precision measurement of electrical charges in an optomechanical system beyond linearized dynamics},\ }\href@noop {} {\bibfield  {journal} {\bibinfo  {journal} {App. Phys. Lett.}\ }\textbf {\bibinfo {volume} {110}} (\bibinfo {year} {2017})}\BibitemShut {NoStop}%
\bibitem [{\citenamefont {Basiri-Esfahani}\ \emph {et~al.}(2019)\citenamefont {Basiri-Esfahani}, \citenamefont {Armin}, \citenamefont {Forstner},\ and\ \citenamefont {Bowen}}]{4}%
  \BibitemOpen
  \bibfield  {author} {\bibinfo {author} {\bibfnamefont {S.}~\bibnamefont {Basiri-Esfahani}}, \bibinfo {author} {\bibfnamefont {A.}~\bibnamefont {Armin}}, \bibinfo {author} {\bibfnamefont {S.}~\bibnamefont {Forstner}},\ and\ \bibinfo {author} {\bibfnamefont {W.~P.}\ \bibnamefont {Bowen}},\ }\bibfield  {title} {\bibinfo {title} {Precision ultrasound sensing on a chip},\ }\href@noop {} {\bibfield  {journal} {\bibinfo  {journal} {Nat. Commun.}\ }\textbf {\bibinfo {volume} {10}},\ \bibinfo {pages} {132} (\bibinfo {year} {2019})}\BibitemShut {NoStop}%
\bibitem [{\citenamefont {Guggenheim}\ \emph {et~al.}(2017)\citenamefont {Guggenheim}, \citenamefont {Li}, \citenamefont {Allen}, \citenamefont {Colchester}, \citenamefont {Noimark}, \citenamefont {Ogunlade}, \citenamefont {Parkin}, \citenamefont {Papakonstantinou}, \citenamefont {Desjardins}, \citenamefont {Zhang} \emph {et~al.}}]{5}%
  \BibitemOpen
  \bibfield  {author} {\bibinfo {author} {\bibfnamefont {J.~A.}\ \bibnamefont {Guggenheim}}, \bibinfo {author} {\bibfnamefont {J.}~\bibnamefont {Li}}, \bibinfo {author} {\bibfnamefont {T.~J.}\ \bibnamefont {Allen}}, \bibinfo {author} {\bibfnamefont {R.~J.}\ \bibnamefont {Colchester}}, \bibinfo {author} {\bibfnamefont {S.}~\bibnamefont {Noimark}}, \bibinfo {author} {\bibfnamefont {O.}~\bibnamefont {Ogunlade}}, \bibinfo {author} {\bibfnamefont {I.~P.}\ \bibnamefont {Parkin}}, \bibinfo {author} {\bibfnamefont {I.}~\bibnamefont {Papakonstantinou}}, \bibinfo {author} {\bibfnamefont {A.~E.}\ \bibnamefont {Desjardins}}, \bibinfo {author} {\bibfnamefont {E.~Z.}\ \bibnamefont {Zhang}}, \emph {et~al.},\ }\bibfield  {title} {\bibinfo {title} {Ultrasensitive plano-concave optical microresonators for ultrasound sensing},\ }\href@noop {} {\bibfield  {journal} {\bibinfo  {journal} {Nat. Photon.}\ }\textbf {\bibinfo {volume} {11}},\ \bibinfo {pages} {714} (\bibinfo {year} {2017})}\BibitemShut {NoStop}%
\bibitem [{\citenamefont {Zhang}\ \emph {et~al.}(2014)\citenamefont {Zhang}, \citenamefont {Ling}, \citenamefont {Chen},\ and\ \citenamefont {Guo}}]{6}%
  \BibitemOpen
  \bibfield  {author} {\bibinfo {author} {\bibfnamefont {C.}~\bibnamefont {Zhang}}, \bibinfo {author} {\bibfnamefont {T.}~\bibnamefont {Ling}}, \bibinfo {author} {\bibfnamefont {S.-L.}\ \bibnamefont {Chen}},\ and\ \bibinfo {author} {\bibfnamefont {L.~J.}\ \bibnamefont {Guo}},\ }\bibfield  {title} {\bibinfo {title} {Ultrabroad bandwidth and highly sensitive optical ultrasonic detector for photoacoustic imaging},\ }\href@noop {} {\bibfield  {journal} {\bibinfo  {journal} {Acs Photonics}\ }\textbf {\bibinfo {volume} {1}},\ \bibinfo {pages} {1093} (\bibinfo {year} {2014})}\BibitemShut {NoStop}%
\bibitem [{\citenamefont {Tian}(2012)}]{7}%
  \BibitemOpen
  \bibfield  {author} {\bibinfo {author} {\bibfnamefont {L.}~\bibnamefont {Tian}},\ }\bibfield  {title} {\bibinfo {title} {Adiabatic state conversion and pulse transmission in optomechanical systems},\ }\href@noop {} {\bibfield  {journal} {\bibinfo  {journal} {Phys. Rev. Lett.}\ }\textbf {\bibinfo {volume} {108}},\ \bibinfo {pages} {153604} (\bibinfo {year} {2012})}\BibitemShut {NoStop}%
\bibitem [{\citenamefont {Tian}(2013)}]{8}%
  \BibitemOpen
  \bibfield  {author} {\bibinfo {author} {\bibfnamefont {L.}~\bibnamefont {Tian}},\ }\bibfield  {title} {\bibinfo {title} {Robust photon entanglement via quantum interference in optomechanical interfaces},\ }\href@noop {} {\bibfield  {journal} {\bibinfo  {journal} {Phys. Rev. Lett.}\ }\textbf {\bibinfo {volume} {110}},\ \bibinfo {pages} {233602} (\bibinfo {year} {2013})}\BibitemShut {NoStop}%
\bibitem [{\citenamefont {Grudinin}\ \emph {et~al.}(2010)\citenamefont {Grudinin}, \citenamefont {Lee}, \citenamefont {Painter},\ and\ \citenamefont {Vahala}}]{9}%
  \BibitemOpen
  \bibfield  {author} {\bibinfo {author} {\bibfnamefont {I.~S.}\ \bibnamefont {Grudinin}}, \bibinfo {author} {\bibfnamefont {H.}~\bibnamefont {Lee}}, \bibinfo {author} {\bibfnamefont {O.}~\bibnamefont {Painter}},\ and\ \bibinfo {author} {\bibfnamefont {K.~J.}\ \bibnamefont {Vahala}},\ }\bibfield  {title} {\bibinfo {title} {Phonon laser action in a tunable two-level system},\ }\href@noop {} {\bibfield  {journal} {\bibinfo  {journal} {Phys. Rev. Lett.}\ }\textbf {\bibinfo {volume} {104}},\ \bibinfo {pages} {083901} (\bibinfo {year} {2010})}\BibitemShut {NoStop}%
\bibitem [{\citenamefont {Zhang}\ \emph {et~al.}(2018)\citenamefont {Zhang}, \citenamefont {Peng}, \citenamefont {Ozdemir}, \citenamefont {Pichler}, \citenamefont {Krimer}, \citenamefont {Zhao}, \citenamefont {Nori}, \citenamefont {Liu}, \citenamefont {Rotter},\ and\ \citenamefont {Yang}}]{10}%
  \BibitemOpen
  \bibfield  {author} {\bibinfo {author} {\bibfnamefont {J.}~\bibnamefont {Zhang}}, \bibinfo {author} {\bibfnamefont {B.}~\bibnamefont {Peng}}, \bibinfo {author} {\bibfnamefont {S.~K.}\ \bibnamefont {Ozdemir}}, \bibinfo {author} {\bibfnamefont {K.}~\bibnamefont {Pichler}}, \bibinfo {author} {\bibfnamefont {D.~O.}\ \bibnamefont {Krimer}}, \bibinfo {author} {\bibfnamefont {G.}~\bibnamefont {Zhao}}, \bibinfo {author} {\bibfnamefont {F.}~\bibnamefont {Nori}}, \bibinfo {author} {\bibfnamefont {Y.-x.}\ \bibnamefont {Liu}}, \bibinfo {author} {\bibfnamefont {S.}~\bibnamefont {Rotter}},\ and\ \bibinfo {author} {\bibfnamefont {L.}~\bibnamefont {Yang}},\ }\bibfield  {title} {\bibinfo {title} {A phonon laser operating at an exceptional point},\ }\href@noop {} {\bibfield  {journal} {\bibinfo  {journal} {Nature Photon.}\ }\textbf {\bibinfo {volume} {12}},\ \bibinfo {pages} {479} (\bibinfo {year} {2018})}\BibitemShut {NoStop}%
\bibitem [{\citenamefont {Jing}\ \emph {et~al.}(2014)\citenamefont {Jing}, \citenamefont {Ozdemir}, \citenamefont {Lu}, \citenamefont {Zhang}, \citenamefont {Yang},\ and\ \citenamefont {Nori}}]{11}%
  \BibitemOpen
  \bibfield  {author} {\bibinfo {author} {\bibfnamefont {H.}~\bibnamefont {Jing}}, \bibinfo {author} {\bibfnamefont {S.}~\bibnamefont {Ozdemir}}, \bibinfo {author} {\bibfnamefont {X.-Y.}\ \bibnamefont {Lu}}, \bibinfo {author} {\bibfnamefont {J.}~\bibnamefont {Zhang}}, \bibinfo {author} {\bibfnamefont {L.}~\bibnamefont {Yang}},\ and\ \bibinfo {author} {\bibfnamefont {F.}~\bibnamefont {Nori}},\ }\bibfield  {title} {\bibinfo {title} {Pt-symmetric phonon laser},\ }\href@noop {} {\bibfield  {journal} {\bibinfo  {journal} {Phys. Rev. Lett.}\ }\textbf {\bibinfo {volume} {113}},\ \bibinfo {pages} {053604} (\bibinfo {year} {2014})}\BibitemShut {NoStop}%
\bibitem [{\citenamefont {Lu}\ \emph {et~al.}(2017)\citenamefont {Lu}, \citenamefont {Ozdemir}, \citenamefont {Kuang}, \citenamefont {Nori},\ and\ \citenamefont {Jing}}]{12}%
  \BibitemOpen
  \bibfield  {author} {\bibinfo {author} {\bibfnamefont {H.}~\bibnamefont {Lu}}, \bibinfo {author} {\bibfnamefont {S.}~\bibnamefont {Ozdemir}}, \bibinfo {author} {\bibfnamefont {L.-M.}\ \bibnamefont {Kuang}}, \bibinfo {author} {\bibfnamefont {F.}~\bibnamefont {Nori}},\ and\ \bibinfo {author} {\bibfnamefont {H.}~\bibnamefont {Jing}},\ }\bibfield  {title} {\bibinfo {title} {Exceptional points in random-defect phonon lasers},\ }\href@noop {} {\bibfield  {journal} {\bibinfo  {journal} {Phys. Rev. Appl.}\ }\textbf {\bibinfo {volume} {8}},\ \bibinfo {pages} {044020} (\bibinfo {year} {2017})}\BibitemShut {NoStop}%
\bibitem [{\citenamefont {Mukhamedyanov}\ \emph {et~al.}(2023)\citenamefont {Mukhamedyanov}, \citenamefont {Zyablovsky},\ and\ \citenamefont {Andrianov}}]{subthreshold}%
  \BibitemOpen
  \bibfield  {author} {\bibinfo {author} {\bibfnamefont {A.}~\bibnamefont {Mukhamedyanov}}, \bibinfo {author} {\bibfnamefont {A.~A.}\ \bibnamefont {Zyablovsky}},\ and\ \bibinfo {author} {\bibfnamefont {E.~S.}\ \bibnamefont {Andrianov}},\ }\bibfield  {title} {\bibinfo {title} {Subthreshold phonon generation in an optomechanical system with an exceptional point},\ }\href@noop {} {\bibfield  {journal} {\bibinfo  {journal} {Opt. Lett.}\ }\textbf {\bibinfo {volume} {48}},\ \bibinfo {pages} {1822} (\bibinfo {year} {2023})}\BibitemShut {NoStop}%
\bibitem [{\citenamefont {Weis}\ \emph {et~al.}(2010)\citenamefont {Weis}, \citenamefont {Rivi{\`e}re}, \citenamefont {Del{\'e}glise}, \citenamefont {Gavartin}, \citenamefont {Arcizet}, \citenamefont {Schliesser},\ and\ \citenamefont {Kippenberg}}]{14}%
  \BibitemOpen
  \bibfield  {author} {\bibinfo {author} {\bibfnamefont {S.}~\bibnamefont {Weis}}, \bibinfo {author} {\bibfnamefont {R.}~\bibnamefont {Rivi{\`e}re}}, \bibinfo {author} {\bibfnamefont {S.}~\bibnamefont {Del{\'e}glise}}, \bibinfo {author} {\bibfnamefont {E.}~\bibnamefont {Gavartin}}, \bibinfo {author} {\bibfnamefont {O.}~\bibnamefont {Arcizet}}, \bibinfo {author} {\bibfnamefont {A.}~\bibnamefont {Schliesser}},\ and\ \bibinfo {author} {\bibfnamefont {T.~J.}\ \bibnamefont {Kippenberg}},\ }\bibfield  {title} {\bibinfo {title} {Optomechanically induced transparency},\ }\href@noop {} {\bibfield  {journal} {\bibinfo  {journal} {science}\ }\textbf {\bibinfo {volume} {330}},\ \bibinfo {pages} {1520} (\bibinfo {year} {2010})}\BibitemShut {NoStop}%
\bibitem [{\citenamefont {Xiong}\ and\ \citenamefont {Wu}(2018)}]{15}%
  \BibitemOpen
  \bibfield  {author} {\bibinfo {author} {\bibfnamefont {H.}~\bibnamefont {Xiong}}\ and\ \bibinfo {author} {\bibfnamefont {Y.}~\bibnamefont {Wu}},\ }\bibfield  {title} {\bibinfo {title} {Fundamentals and applications of optomechanically induced transparency},\ }\href@noop {} {\bibfield  {journal} {\bibinfo  {journal} {Appl. Phys. Rev.}\ }\textbf {\bibinfo {volume} {5}} (\bibinfo {year} {2018})}\BibitemShut {NoStop}%
\bibitem [{\citenamefont {Mukhamedyanov}\ \emph {et~al.}(2024)\citenamefont {Mukhamedyanov}, \citenamefont {Zyablovsky},\ and\ \citenamefont {Andrianov}}]{hard_exc}%
  \BibitemOpen
  \bibfield  {author} {\bibinfo {author} {\bibfnamefont {A.}~\bibnamefont {Mukhamedyanov}}, \bibinfo {author} {\bibfnamefont {A.~A.}\ \bibnamefont {Zyablovsky}},\ and\ \bibinfo {author} {\bibfnamefont {E.~S.}\ \bibnamefont {Andrianov}},\ }\bibfield  {title} {\bibinfo {title} {Hard excitation mode of a system with optomechanical instability},\ }\href@noop {} {\bibfield  {journal} {\bibinfo  {journal} {Opt. Lett.}\ }\textbf {\bibinfo {volume} {49}},\ \bibinfo {pages} {782} (\bibinfo {year} {2024})}\BibitemShut {NoStop}%
\bibitem [{\citenamefont {Melentiev}\ \emph {et~al.}(2017)\citenamefont {Melentiev}, \citenamefont {Kalmykov}, \citenamefont {Gritchenko}, \citenamefont {Afanasiev}, \citenamefont {Balykin}, \citenamefont {Baburin}, \citenamefont {Ryzhova}, \citenamefont {Filippov}, \citenamefont {Rodionov}, \citenamefont {Nechepurenko} \emph {et~al.}}]{16}%
  \BibitemOpen
  \bibfield  {author} {\bibinfo {author} {\bibfnamefont {P.}~\bibnamefont {Melentiev}}, \bibinfo {author} {\bibfnamefont {A.}~\bibnamefont {Kalmykov}}, \bibinfo {author} {\bibfnamefont {A.}~\bibnamefont {Gritchenko}}, \bibinfo {author} {\bibfnamefont {A.}~\bibnamefont {Afanasiev}}, \bibinfo {author} {\bibfnamefont {V.}~\bibnamefont {Balykin}}, \bibinfo {author} {\bibfnamefont {A.}~\bibnamefont {Baburin}}, \bibinfo {author} {\bibfnamefont {E.}~\bibnamefont {Ryzhova}}, \bibinfo {author} {\bibfnamefont {I.}~\bibnamefont {Filippov}}, \bibinfo {author} {\bibfnamefont {I.}~\bibnamefont {Rodionov}}, \bibinfo {author} {\bibfnamefont {I.}~\bibnamefont {Nechepurenko}}, \emph {et~al.},\ }\bibfield  {title} {\bibinfo {title} {Plasmonic nanolaser for intracavity spectroscopy and sensorics},\ }\href@noop {} {\bibfield  {journal} {\bibinfo  {journal} {App. Phys. Lett.}\ }\textbf {\bibinfo {volume} {111}} (\bibinfo {year} {2017})}\BibitemShut {NoStop}%
\bibitem [{\citenamefont {Ma}\ \emph {et~al.}(2014)\citenamefont {Ma}, \citenamefont {Ota}, \citenamefont {Li}, \citenamefont {Yang},\ and\ \citenamefont {Zhang}}]{17}%
  \BibitemOpen
  \bibfield  {author} {\bibinfo {author} {\bibfnamefont {R.-M.}\ \bibnamefont {Ma}}, \bibinfo {author} {\bibfnamefont {S.}~\bibnamefont {Ota}}, \bibinfo {author} {\bibfnamefont {Y.}~\bibnamefont {Li}}, \bibinfo {author} {\bibfnamefont {S.}~\bibnamefont {Yang}},\ and\ \bibinfo {author} {\bibfnamefont {X.}~\bibnamefont {Zhang}},\ }\bibfield  {title} {\bibinfo {title} {Explosives detection in a lasing plasmon nanocavity},\ }\href@noop {} {\bibfield  {journal} {\bibinfo  {journal} {Nat. nanotechnology}\ }\textbf {\bibinfo {volume} {9}},\ \bibinfo {pages} {600} (\bibinfo {year} {2014})}\BibitemShut {NoStop}%
\bibitem [{\citenamefont {Zasedatelev}\ \emph {et~al.}(2019)\citenamefont {Zasedatelev}, \citenamefont {Baranikov}, \citenamefont {Urbonas}, \citenamefont {Scafirimuto}, \citenamefont {Scherf}, \citenamefont {St{\"o}ferle}, \citenamefont {Mahrt},\ and\ \citenamefont {Lagoudakis}}]{19}%
  \BibitemOpen
  \bibfield  {author} {\bibinfo {author} {\bibfnamefont {A.~V.}\ \bibnamefont {Zasedatelev}}, \bibinfo {author} {\bibfnamefont {A.~V.}\ \bibnamefont {Baranikov}}, \bibinfo {author} {\bibfnamefont {D.}~\bibnamefont {Urbonas}}, \bibinfo {author} {\bibfnamefont {F.}~\bibnamefont {Scafirimuto}}, \bibinfo {author} {\bibfnamefont {U.}~\bibnamefont {Scherf}}, \bibinfo {author} {\bibfnamefont {T.}~\bibnamefont {St{\"o}ferle}}, \bibinfo {author} {\bibfnamefont {R.~F.}\ \bibnamefont {Mahrt}},\ and\ \bibinfo {author} {\bibfnamefont {P.~G.}\ \bibnamefont {Lagoudakis}},\ }\bibfield  {title} {\bibinfo {title} {A room-temperature organic polariton transistor},\ }\href@noop {} {\bibfield  {journal} {\bibinfo  {journal} {Nat. Phot.}\ }\textbf {\bibinfo {volume} {13}},\ \bibinfo {pages} {378} (\bibinfo {year} {2019})}\BibitemShut {NoStop}%
\bibitem [{\citenamefont {Zasedatelev}\ \emph {et~al.}(2021)\citenamefont {Zasedatelev}, \citenamefont {Baranikov}, \citenamefont {Sannikov}, \citenamefont {Urbonas}, \citenamefont {Scafirimuto}, \citenamefont {Shishkov}, \citenamefont {Andrianov}, \citenamefont {Lozovik}, \citenamefont {Scherf}, \citenamefont {St{\"o}ferle} \emph {et~al.}}]{20}%
  \BibitemOpen
  \bibfield  {author} {\bibinfo {author} {\bibfnamefont {A.~V.}\ \bibnamefont {Zasedatelev}}, \bibinfo {author} {\bibfnamefont {A.~V.}\ \bibnamefont {Baranikov}}, \bibinfo {author} {\bibfnamefont {D.}~\bibnamefont {Sannikov}}, \bibinfo {author} {\bibfnamefont {D.}~\bibnamefont {Urbonas}}, \bibinfo {author} {\bibfnamefont {F.}~\bibnamefont {Scafirimuto}}, \bibinfo {author} {\bibfnamefont {V.~Y.}\ \bibnamefont {Shishkov}}, \bibinfo {author} {\bibfnamefont {E.~S.}\ \bibnamefont {Andrianov}}, \bibinfo {author} {\bibfnamefont {Y.~E.}\ \bibnamefont {Lozovik}}, \bibinfo {author} {\bibfnamefont {U.}~\bibnamefont {Scherf}}, \bibinfo {author} {\bibfnamefont {T.}~\bibnamefont {St{\"o}ferle}}, \emph {et~al.},\ }\bibfield  {title} {\bibinfo {title} {Single-photon nonlinearity at room temperature},\ }\href@noop {} {\bibfield  {journal} {\bibinfo  {journal} {Nature}\ }\textbf {\bibinfo {volume} {597}},\ \bibinfo {pages} {493} (\bibinfo {year} {2021})}\BibitemShut {NoStop}%
\bibitem [{\citenamefont {Altug}\ \emph {et~al.}(2006)\citenamefont {Altug}, \citenamefont {Englund},\ and\ \citenamefont {Vu{\v{c}}kovi{\'c}}}]{21}%
  \BibitemOpen
  \bibfield  {author} {\bibinfo {author} {\bibfnamefont {H.}~\bibnamefont {Altug}}, \bibinfo {author} {\bibfnamefont {D.}~\bibnamefont {Englund}},\ and\ \bibinfo {author} {\bibfnamefont {J.}~\bibnamefont {Vu{\v{c}}kovi{\'c}}},\ }\bibfield  {title} {\bibinfo {title} {Ultrafast photonic crystal nanocavity laser},\ }\href@noop {} {\bibfield  {journal} {\bibinfo  {journal} {Nature physics}\ }\textbf {\bibinfo {volume} {2}},\ \bibinfo {pages} {484} (\bibinfo {year} {2006})}\BibitemShut {NoStop}%
\bibitem [{\citenamefont {Nefedkin}\ \emph {et~al.}(2019)\citenamefont {Nefedkin}, \citenamefont {Zyablovsky}, \citenamefont {Andrianov}, \citenamefont {Pukhov},\ and\ \citenamefont {Vinogradov}}]{22}%
  \BibitemOpen
  \bibfield  {author} {\bibinfo {author} {\bibfnamefont {N.}~\bibnamefont {Nefedkin}}, \bibinfo {author} {\bibfnamefont {A.}~\bibnamefont {Zyablovsky}}, \bibinfo {author} {\bibfnamefont {E.}~\bibnamefont {Andrianov}}, \bibinfo {author} {\bibfnamefont {A.}~\bibnamefont {Pukhov}},\ and\ \bibinfo {author} {\bibfnamefont {A.}~\bibnamefont {Vinogradov}},\ }\bibfield  {title} {\bibinfo {title} {Response time of a plasmonic distributed feedback laser in a large-signal modulation regime},\ }\href@noop {} {\bibfield  {journal} {\bibinfo  {journal} {Physical Review Applied}\ }\textbf {\bibinfo {volume} {11}},\ \bibinfo {pages} {054067} (\bibinfo {year} {2019})}\BibitemShut {NoStop}%
\bibitem [{\citenamefont {Allen}\ and\ \citenamefont {Eberly}(2012)}]{allen}%
  \BibitemOpen
  \bibfield  {author} {\bibinfo {author} {\bibfnamefont {L.}~\bibnamefont {Allen}}\ and\ \bibinfo {author} {\bibfnamefont {J.~H.}\ \bibnamefont {Eberly}},\ }\href@noop {} {\emph {\bibinfo {title} {Optical resonance and two-level atoms}}}\ (\bibinfo  {publisher} {Courier Corporation},\ \bibinfo {year} {2012})\BibitemShut {NoStop}%
\bibitem [{\citenamefont {Carmichael}(2009)}]{carmichael}%
  \BibitemOpen
  \bibfield  {author} {\bibinfo {author} {\bibfnamefont {H.}~\bibnamefont {Carmichael}},\ }\href@noop {} {\emph {\bibinfo {title} {An open systems approach to quantum optics: lectures presented at the Universit{\'e} Libre de Bruxelles, October 28 to November 4, 1991}}},\ Vol.~\bibinfo {volume} {18}\ (\bibinfo  {publisher} {Springer Science \& Business Media},\ \bibinfo {year} {2009})\BibitemShut {NoStop}%
\bibitem [{\citenamefont {Gardiner}\ and\ \citenamefont {Zoller}(2004)}]{gardiner}%
  \BibitemOpen
  \bibfield  {author} {\bibinfo {author} {\bibfnamefont {C.}~\bibnamefont {Gardiner}}\ and\ \bibinfo {author} {\bibfnamefont {P.}~\bibnamefont {Zoller}},\ }\href@noop {} {\emph {\bibinfo {title} {Quantum noise: a handbook of Markovian and non-Markovian quantum stochastic methods with applications to quantum optics}}}\ (\bibinfo  {publisher} {Springer Science \& Business Media},\ \bibinfo {year} {2004})\BibitemShut {NoStop}%
\bibitem [{\citenamefont {Scully}\ and\ \citenamefont {Zubairy}(1997)}]{39}%
  \BibitemOpen
  \bibfield  {author} {\bibinfo {author} {\bibfnamefont {M.~O.}\ \bibnamefont {Scully}}\ and\ \bibinfo {author} {\bibfnamefont {M.~S.}\ \bibnamefont {Zubairy}},\ }\href@noop {} {\emph {\bibinfo {title} {Quantum optics}}}\ (\bibinfo  {publisher} {Cambridge university press},\ \bibinfo {year} {1997})\BibitemShut {NoStop}%
\end{thebibliography}%

\end{document}